\documentclass{article}
\usepackage[dvips]{graphicx}
\usepackage{rotating}
\begin{document}
\bibliographystyle{plain}

\begin{center}
\Large An Inverse Equilibrium Maximum Entropy Algorithm Applied To Proteins\\

\normalsize John P. Donohue\\
Department of Physics, University of California, Santa Cruz, California 95064 \\
\date{\today}

\end{center}



\begin{abstract}
A computational method is developed to work on an inverse equilibrium problem
with an interest towards applications with protein folding. In general, we
are given a set of equilibrium configurations, and want to derive
the most likely  potential function that results in 
these configurations. The method is applied to polymer simulations and 
a simple model of proteins using protein structures obtained from the 
Protein Data Bank (http://www.rcsb.org/pdb)
The resulting energy function is tested on a few decoy sets 
with limited success.
\end{abstract}

\section*{Introduction}
The Protein Data Bank (http://www.rcsb.org/pdb) \cite{pdb} has about
19,000 protein structures solved by several
methods. There are many algorithms that use this information
to derive understanding about protein interactions. Our method
is based on using these equilibrium configurations and a 
maximum entropy algorithm to derive information about
physical energy functions that could be used to approximate
protein interactions.

\section*{Our Method}
The following is based largely on an algorithm originally developed by 
reference \cite{det2}.
The method is based on the following assumptions. The system is assumed 
to be in thermodynamic equilibrium.  For proteins, this was shown to 
be a good assumption by 
Anfinsen \cite{Anfinsen}.  We also assume 
the energy can be written as a sum of terms which are products
of parameters and functions of the configuration.
$ E(\Gamma,\vec{P}) = \sum_{i} p_{i} * h_{i}(\Gamma) = \vec{P} \cdot \vec{H}$
where $\Gamma$ represents the configuration of the system(s).
$\vec{P} = \{p_{i}\}$ represents the set of parameters to be derived.

The probability of a configuration, given parameters, is given by the 
Boltzmann distribution
$ Prob(\Gamma | \vec{P}) = e^{-E(\Gamma , \vec{P})/kT} / Z = e^{(-E(\Gamma , \vec{P}) + F(\vec{P})) / kT} $ ,where
$ Z(\vec{P}) = \sum_{\Gamma}Exp(-\beta E(\Gamma,\vec{P}))$ and
$ F(\vec{P}) =  -kT ln(Z(\vec{P}))$.

If we are given the exact equilibrium conformation, $\Gamma^*$ ,
the maximum likelihood of parameter values are those values for which the 
probability, $Prob(\Gamma^* | \vec{P})$ is a maximum wrt $\vec{P}$.
Maximizing an exponential corresponds
to maximizing the argument (ignoring the multiplicative constant $\beta$), 
$ -E(\Gamma^*,\vec{P}) + F(\vec{P}) = Q(\vec{P}) $.
This also corresponds to extremizing the entropy $TS = E - F $.

Our method is basically the multi-dimensional form of 
Newton's method for optimizing functions.
Maximizing $Q(\vec{P})$, Newton's Method is
\begin{equation}\vec{P}^{k+1} = \vec{P}^{k} - D^2(Q(\vec{P}^{k}))^{-1} \cdot D(Q(\vec{P}^{k}))\end{equation}
where $(D^2)^{-1}$ represents the inverse Hessian matrix 
and D represents the gradient.
In practice this is modified slightly,
\begin{equation}\vec{P}^{k+1} = \vec{P}^{k} + \epsilon(\Delta \vec{P})\end{equation}
where the use of $ \epsilon < 1 $ corresponds to the "Damped Newton's Method".

Using statistical mechanical definitions
\begin{equation} \frac{\partial (-E(\Gamma^*)) }{\partial p_i} = -h_i^*  
\hspace{0.5in}\frac{\partial^2 (-E(\Gamma^*)) }{\partial p_i \partial p_j} = 0 \end{equation} 
\begin{equation} \frac{\partial F}{\partial P_i}|_{P^k} = -kT (1/Z) \sum_{\Gamma} -\beta h_i e^{-\beta E(\Gamma, P)} = <h_i> \end{equation}
\begin{equation} \frac{\partial^2 F} {\partial p_i \partial p_j} = \beta ( <h_i> <h_j> - <h_i h_j> )\end{equation}
Maximizing $ Q = -E + F $ wrt $\vec{P}$
 leads to the following.
\begin{equation} D(Q)_i = -h_i^* + <h_i> \end{equation} 
\begin{equation} D^2(Q)_{i,j} = \beta ( <h_i> <h_j> - <h_i h_j> ) = - \beta Cov(h_i,h_j) \end{equation}
Resulting in the following iterative
equation where $VCM(\vec{H})$ is the variance-covariance matrix of $\vec{H}$
\begin{equation} \Delta \vec{P} = kT*VCM(\vec{H})^{-1} \bullet (<\vec{H}> - \vec{H^*} )\end{equation}
The method is easily generalized to a distribution of equilibrium configurations.
\begin{equation} \Delta \vec{P} =  kT*VCM(\vec{H})^{-1} \bullet (<\vec{H}> - <\vec{H}>_{Prob(\Gamma)}) \end{equation}
$<...>$ represents a Boltzmann average and $<...>_{Prob(\Gamma)}$ represents an average over
the given distribution.

Assuming the least prior information, the iteration starts with 
all parameters set to zero.
This would not allow any useful MC evolution at all. 
The energy would be zero for 
any Monte Carlo move, thus not preferring any particular move.
The energy is modified for the first few iterations with the addition
of a clamping term.
\begin{equation} E_{clamp} = \sum P_{clamp} * (r - r^*)^2 \end{equation}
This term  makes the given conformation a minimum of the 
energy. Since this conformation is an equilibrium conformation, this
seems to be a good approximation. Once the parameter values are sufficiently
away from zero, this term is set to zero.
If a distribution of conformations is given, this distribution can be used
as the clamping terms.

\subsection*{Computation}

The basic algorithm is as follows.
\begin{itemize}
\item
Given $P^{k}$, Monte Carlo simulations and averaging are used to find
$<h_j>$ and $<h_i h_j>$. 
\item
This leads to a matrix equation.
$ \Delta \vec{P} =  kT*VCM(\vec{H})^{-1} \bullet (<\vec{H}> - \vec{H}^*) $
\item
Solve this for $\Delta \vec{P}$
\item 
$\vec{P}^{k+1} = \vec{P}^{k} + \epsilon(\Delta \vec{P})$
\item 
Repeat until $\vec{P}$ converges.
\end{itemize}

\subsection*{Computation: Convergence time}

Assuming a system of M proteins, each containing 
N particles with P parameters.
The Monte Carlo time required for useful averaging scales at least as $N$.
The pair-wise energy calculation in each Monte Carlo step scales as $N^2$.
Although for some models a radius cutoff and neighbor list were used, 
so this effect in practice is less than indicated.
Solving a P by P matrix equation is a $P^3$ operation.
Computation time for this method scales as \begin{equation} \alpha M *  N^3  +  \beta P^3\end{equation}

\section*{Simulation}

A Monte Carlo evolution was done on several systems.
Four energy terms were used.

\begin{equation} \pm \frac{a}{|r_i - r_j|} + \frac{b}{(r_i - r_j)^{12}} +
 c*|r_i - r_{i+1}| + d*(r_i - r_{i+1})^2 \end{equation}

Non-bonded interactions were included similar to an electrostatic 
attraction and van der Waals repulsion.
The sequence of positive and negative charges was randomly chosen.
A covalent-type bond between one particle and the next in the sequence
was given a quadratic form.
For the simulation described below, the following parameters
were used $ a=2 , b=4 , c=-4 , d=1$.
The results of the algorithm applied to this system are shown in figure 1 
and table 1.

\begin{figure}[h]
\includegraphics[angle=-90,width=5in]{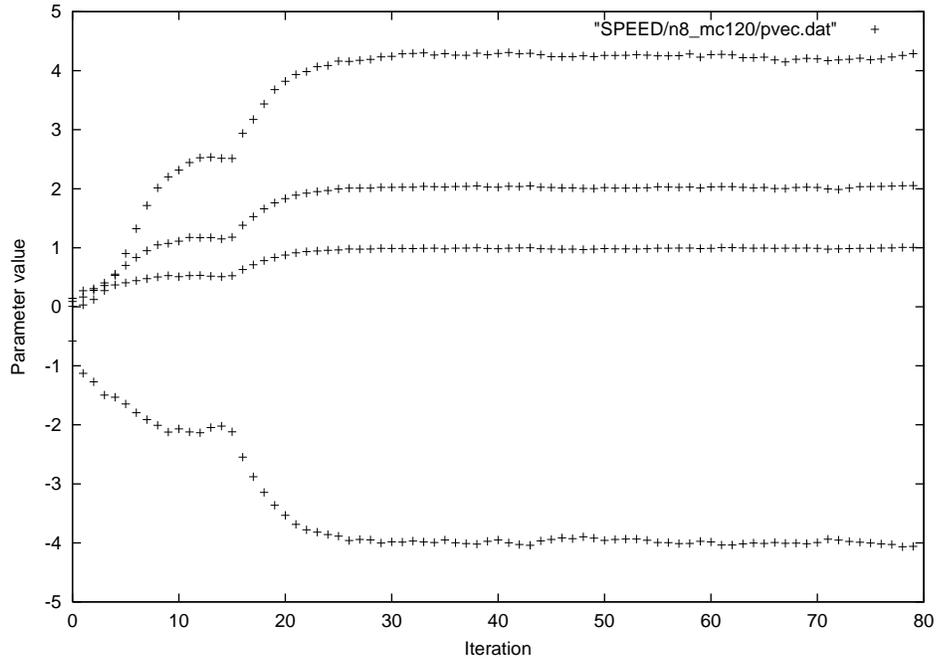}
\caption{
A collection of 99, 8 particle proteins was constructed
with $ a=2 , b=4 , c=-4 , d=1, kt=0.25$ The algorithm used $ \epsilon=0.5 $ and
120 MC moves per particle per iteration.
Clamping was turned off at time=16
Using a computer with dual P3 450-667Mhz this took 1.5 hrs}
\end{figure}

\begin{table}[h]
\begin{tabular}{l|r}
Correct Value & Derived value \\ 
\hline
2 &  $2.01  \pm 0.01 $ \\ 
\hline
4 &  $4.36  \pm 0.03$ \\ 
\hline
-4 & $ -3.88  \pm 0.03$ \\ 
\hline
1 &  $0.97  \pm 0.01$ \\ 
\hline
\end{tabular}
\caption{Parameters derived from equilibrium configurations compare well
with correct values. }
\end{table}

\clearpage

\subsection*{Protein Model}

Several major simplifications were made to allow convergence of real
protein parameters in a reasonable time.
The united residue approximation is used in the 
following model. Each of the residues is treated as one particle. 
This approximation is commonly used \cite{banavar,paydirt,liwo}.
This greatly simplifies the system, but should contain enough complexity
to describe the system adequately.

Proteins seem to have a rugged free energy surface.
To minimize this effect we use relatively few MC time steps
per iteration. This keeps the protein in the local minimum of
free energy even if parameters are far from correct.

Residues were placed at the $C_{\alpha}$ location, 
covalently bond only with next and previous particles on the chain.
The energy function used for the covalent bonds was a normal distribution 
using the mean and variance derived from the data. 
The noise in the covalent bond parameter seemed to cause large 
perturbations in the convergence for other parameters.
These parameters need to be derived separately.

The first model with consistent convergence used a statistical 
grouping of residues developed by Cieplak etal
\cite{mariban}. The groupings were derived using a simplification of a 
statistically derived interaction  matrix, 
the Miyazawa Jernigan (MJ) matrix \cite{mjmat}. 
This is a simple, consistent grouping which 
decreases the number of parameters.

\begin{itemize}
\item Hydrophobic I (ave hydrophobicity scale value 2.6)
(LFI) Leucine, Isoleucine and Phenylalanine

\item Hydrophobic II (ave HP scale value 1.8 with large variance)
(MVWCY) Methionine, Valine, Tryptophan, Cysteine and Tyrosine

\item Polar I (ave HP scale value 1.15)
(HA) Histidine, Alanine

\item Polar II (ave HP scale value 0.6)
(TGPRQSNED)
Threonine, Glycine, Proline, Arginine, Glutamine, Serine, Asparagine 
Glutamic acid and Aspartic acid

\item Lysine (ave HP scale value 1.9)
(K) Lysine

\end{itemize}

Only one energy term was used corresponding to  van der Waals attraction 
\begin{equation} 4\epsilon[ (\sigma/r)^{12} -  (\sigma/r)^{6}] \end{equation} 
Essentially this is a contact energy function.
$\sigma$ was determined by 
comparing typical volumes and treating the residues as spheres. 
This gives radii from 2.4 - 3.8 Angstrom.
$\sigma$ in the equation corresponds
to where the core repulsion occurs (about 2*radius) so a 
value of 5 was arbitrarily assumed.

In summary, the covalent properties were approximated from the 
mean and variance of bond lengths. The energy function has one term,
a 6-12 combined term. A statistical grouping was used to further reduce
parameters. This grouping and energy function model has 15 parameters.
Results shown below were derived from proteins ranging in 
size from 20 to 400 residues. For consistency, only X-ray data
and only complete proteins containing no extraneous molecules were used.
The training set contained 821 proteins.
All protein
structures were obtained from the Protein Data Bank \cite{pdb}.
Only 20 MC steps per particle were used. 

\subsection*{Protein Models - Results}

Energy function: $ 4\epsilon [ (\sigma/r)^{12} -  (\sigma/r)^{6}] 
$ with $\sigma=5$

\begin{table}[h]
\begin{tabular}{l|r|r|r|r|r}
Group & Hydrophobic I & H II & Polar I & P II & Lysine\\
\hline
H I  & 0.039 & 0.033 & 0.039 & 0 & 0.015 \\
\hline
H II & & 0.042 & 0.042 & 0 & 0.038  \\
\hline
P I  & & & 0.033 & 0 & 0.036  \\
\hline
P II & & & & 0 & 0.018 \\
\hline
Lysine & & & & & 0.020 \\
\end{tabular}
\caption{Attractive contact energy. Units are kT, $\sigma=5$ }
\end{table}

The zeros were artificially created, 
as the algorithm can not handle these parameters going negative. 
$(\epsilon \geq 0) $ This would cause the MC to diverge.
Despite this limitation, convergence was achieved.
These results imply the least hydrophobic 
(most polar) group essentially has no non-bonded interactions. 
This is very similar to the HP model of polymers where
hydrophobic collapse is modeled as HH atraction 
and other interactions (HP and PP) are ignored.
This took 14 days on a Dual PIII 450MHz.

\begin{figure}[h]
\includegraphics[angle=-90,width=4in]{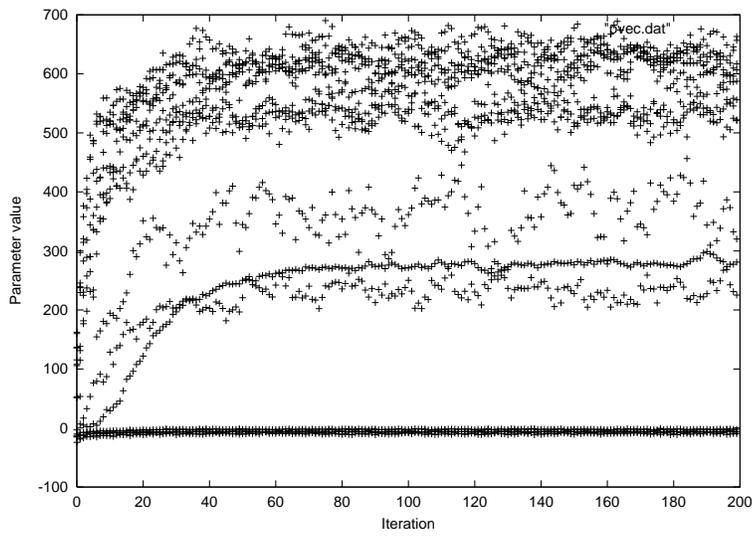}
\caption{Parameters vs Iteration time showing a distribution of variances
in derived parameters}
\end{figure}

\clearpage

\subsection*{Protein Models - Model Evaluation}

This energy function was applied to several decoy sets. Despite the 
simplicity of the energy function,
results were mixed with several very encouraging successes.
All decoy sets were obtained from http://dd.stanford.edu/

\subsection*{4 State Reduced Decoy Set \cite{parklev96}}

\begin{table}[h]
\begin{tabular}{l|r|r|r|r|r|r}
Protein & Rank & Correct structure & ave energy & $\Delta/\sigma$ & present in data set? \\
\hline
1ctf& 1/631& 16.9 & 19.3 & 1.9 & n\\
\hline
1r69& 1/676 & 8.7 & 9.9 & 1.5 & y \\
\hline
1sn3& 10/661 & 16.3 & 18.1 & 1.3 & n\\
\hline
2cro& 1/675& 14.4 & 15.8 & 1.3 & n\\
\hline
3icb& 25/654& 14.6 &15.5&0.9 & y\\
\hline
4pti& 574/688& 12.9 &12.3& -0.7 & y\\
\hline
4rxn& 347/678& 43.2 &52.8&0.5 & y \\
\end{tabular}
\caption{Ranking of correct structure energy using our energy function using the 4state reduced decoy set from Park and Levitt, 1996 \cite{parklev96} }
\end{table}

\subsection*{Local Minima Decoy Set (lmds) \cite{kesar}}

Similar analysis was done for the lmds decoy set from 
C Kesar and M Levitt, 1999.  
These results were not very encouraging. 
The correct proteins were the worst scores in almost 
all cases and by a large amount. 
Currently the main weakness is due to the difference in derivation for the 
covalent bonds. This decoy set was created using a minimization of a 
backbone torsional energy function, hence was very different 
from our function.
Essentially, our successful decoy set predictions are based on
only non-bonded interaction calculations.

\subsection*{Fisa Casp3 Decoy Set \cite{bakdecoy}}

\begin{table}[h]
\begin{tabular}{l|r|r|r|r|r|r}
Protein & Rank & Correct structure & ave energy & $\Delta/\sigma$ & present in data set? \\
\hline
1bl0& 537/972& 17.6 & 18.9 & 0.2 & n \\
\hline
1eh2& 1/2414 & 24.0 & 26.2 & 0.5 & n \\
\hline
1bg8-A& 1/1201 & 16.3 & 17.1 & 0.3 & n \\
\hline
1jwe & 1/1408 & 18.7 & 21.9  & 0.7 & n \\
\hline
smd3 & 226/1201 & 12.4 & 13.3 & 0.5 & n \\
\hline
\end{tabular}
\caption{Ranking of correct structure energy using our energy function using fisa casp3 decoy set from  Simons KT,et al, 1997 \cite{bakdecoy}. None were present in the data used in the derivation. }
\end{table}

\clearpage

\subsection*{Protein Models - Discussion}

The energy function used is extremely simple. Despite this,
ranking of the native protein for 
some of the decoys was very encouraging. 
The poor performers probably need more complex energy functions.
Correlation between RMDS and energy was investigated, but no simple
relationship was found. For the best performers, the native energy was
typically isolated at the lowest energy with most decoys 
concentrated a distinct difference in energy away.

The algorithm has potential, but several problems must be overcome.
Complexity has not been handled
very well and may be required for applicability.
Inclusion of Coulomb, torsional, angular and backbone potential terms
are required for realistic models.
All atom and explicit solvent are further steps that can be taken.
Terms of smaller magnitude are dominated by effects of terms with
larger magnitude and have to be separately derived.
Similarly, terms with smaller frequency of occurrence are dominated
by more frequent terms.
This separate derivation of terms can be organized and iterated
consistently.

\subsection*{Acknowledgements}

J Deutsch, Lik Wee Lee, Leif Poorman, Stefan Meyer, TJ Cox, B Allgood

\bibliography{mybib}

\end{document}